\documentclass[%
 reprint,
superscriptaddress,
 amsmath,amssymb,
 aps,twocolumn
]{revtex4-2}
\usepackage{float}
\usepackage{graphicx}
\usepackage{subcaption}
\usepackage{dcolumn}
\usepackage{bm}
\usepackage{hyperref}
\usepackage{natbib}
\usepackage{caption}
\usepackage{amsmath}
\usepackage{array}
\usepackage{makecell}
\usepackage{dblfloatfix}
\usepackage{multirow}
\usepackage{afterpage}
\usepackage{svg}
\usepackage{tikz}
\usepackage{cellspace}

\begin{document}

\preprint{APS/123-QED}

\title{The Importance of Pure Dephasing in the Optical Response of Excitons in High-quality van der Waals Heterostructures\\}

\author{Anabel Atash Kahlon}
\affiliation{School of Electrical Engineering, Faculty of Engineering, Tel Aviv University, Tel Aviv 6997801, Israel}
\author{Matan Meshulam}
\affiliation{School of Electrical Engineering, Faculty of Engineering, Tel Aviv University, Tel Aviv 6997801, Israel}
\affiliation{School of Physics and Astronomy, Tel Aviv University, Tel Aviv, 6997801, Israel}
\author{Tomer Eini}
\affiliation{School of Electrical Engineering, Faculty of Engineering, Tel Aviv University, Tel Aviv 6997801, Israel}
\author{Thomas Poirier}
\affiliation{Tim Taylor Department of Chemical Engineering, Kansas State University, Manhattan, KS, USA}
\author{James H. Edgar}
\affiliation{Tim Taylor Department of Chemical Engineering, Kansas State University, Manhattan, KS, USA}
\author{Seth Ariel Tongay}
\affiliation{Materials Science and Engineering, School for Engineering of Matter, Transport and Energy,
Arizona State University, Tempe, Arizona 85287, USA}
\author{Itai Epstein}
\email{itaieps@tauex.tau.ac.il}
\affiliation{School of Electrical Engineering, Faculty of Engineering, Tel Aviv University, Tel Aviv 6997801, Israel}
\affiliation{Center for Light-Matter Interaction, Tel Aviv University, Tel Aviv 6997801, Israel}
\affiliation{QuanTAU, Quantum Science and Technology Center, Tel Aviv University, Tel Aviv 6997801, Israel}

\begin{abstract}
Excitons in monolayer transition metal dichalcogenides (TMDs) dominate their optical response due to exceptionally large binding energies arising from their two-dimensional nature. Several theoretical models have been proposed to describe this excitonic behavior, however, it remains unclear which model most accurately captures the underlying physical properties of the response. In this work, we experimentally measure the optical response of high-quality monolayer TMD heterostructures and compare the results with the different theoretical models to address this uncertainty. We find that in high-quality heterostructures, quantum mechanical interactions in the form of pure dephasing plays a dominant role, which has been challenging to isolate experimentally in previous studies. Furthermore, accounting for an additional decay rate to the commonly used radiative and non-radiative rates is found to be important for the accurate description of the excitonic response. These findings establish a robust framework for understanding and predicting the optical properties of TMD-based heterostructures, crucial for both fundamental research and optoelectronic applications.
\end{abstract}

\maketitle

\section{Introduction}
Monolayer transition metal dichalcogenides (TMDs) have attracted significant scientific attention owing to their unique properties, such as strong spin-orbit coupling and direct bandgap , together with their extraordinary optical properties, such as valley-dependent optical selection rules and high photoluminescence quantum yield \cite{Xiao2012CoupledDichalcogenides, Mak2010AtomicallySemiconductor, Splendiani2010EmergingMoS2, Mak2012ControlHelicity, Sallen2012RobustExcitation, KimInhibitedSemiconductors, Amani2015Near-unitysub2/sub}. The two-dimensional nature of TMDs leads to reduced dielectric screening and enhanced Coulomb interactions, resulting in excitons with exceptionally large binding energies \cite{Chernikov2014ExcitonWS2, Wang2018Colloquium:Dichalcogenides} that dominate their optical response. Consequently, excitons in TMDs are robust and interact strongly with light, even under ambient conditions \cite{Wang2018Colloquium:Dichalcogenides, Mueller2018ExcitonSemiconductors}. These characteristics make TMDs promising candidates for novel optoelectronic and valleytronic applications \cite{Mueller2018ExcitonSemiconductors, Lynch2022ExcitonOptics, Lynch2022ExcitonOpticsb}.\par

One of the most important attributes of the excitonic optical response is the exciton linewidth, comprising both homogeneous and inhomogeneous contributions. The homogeneous broadening arises from the population relaxation that includes the radiative and non-radiative lifetime, and interactions such as exciton-phonon and exciton-exciton scattering \cite{Cadiz2017ExcitonicHeterostructures, Moody2015IntrinsicDichalcogenides, Selig2016ExcitonicDichalcogenides, Moody2016ExcitonInvited, Robert2016ExcitonMonolayers}. Inhomogeneous broadening, in contrast, originates from local variations in the exciton's environment, such as external dielectric environment, strain, impurities, and defects \cite{Rhodes2019DisorderMaterials, Raja2017CoulombMaterials, Aslan2018StrainWSe2, Aslan2020StrainedCoupling, Ajayi2017ApproachingMonolayers, Liu2016EngineeringEffects}.

For monolayers commonly placed on $SiO_2$ the typical exciton linewidth have been measured to be on the order of $~10 meV$ at $T=4K$ \cite{Cadiz2016WellMonolayers, Li2014MeasurementE2}. However, when encapsulated in hexagonal boron nitride (hBN), they exhibit narrower linewidths on the order of $2-5 meV$ (at $T=4K$), and have been shown to reach the exciton's homogeneous linewidth \cite{Cadiz2017ExcitonicHeterostructures, Ajayi2017ApproachingMonolayers, Epstein2020Near-unityCavity, Shree2019HighDeposition}. This stems from the fact that hBN-encapsulation effectively isolates the monolayers from environmental disorders, which significantly reduces electrostatic effects and possible stress on the TMD monolayer. Owing to the superior optical properties achieved under these conditions, in this work we will focus exclusively on hBN-encapsulated TMD heterostructure.\par
 
 Various studies have previously focused on modeling the optical response of TMD excitons by employing both theoretical approaches and experimental measurements in the form of photoluminescence and reflection spectroscopy \cite{Epstein2020HighlySemiconductors, Scuri2018LargeNitride, Rogers2020CoherentExcitons, Horng2020PerfectCrystal, Li2021RefractiveDiselenide, Horng2019EngineeringSemiconductors, Fang2019ControlHeterostructures, Robert2018OpticalHeterostructures, Goncalves2018Plasmon-excitonInterfaces, Geisler2019Single-CrystallineTemperature, Back2018RealizationMoSe2, Shree2021GuideSemiconductors}. From these, several models present a distinct perspective on the governing exciton physics, emphasizing different physical constituents and employing unique mathematical approaches to describe the optical response \cite{Scuri2018LargeNitride, Horng2020PerfectCrystal, Epstein2020Near-unityCavity, Rogers2020CoherentExcitons, Li2021RefractiveDiselenide, Goncalves2018Plasmon-excitonInterfaces, Geisler2019Single-CrystallineTemperature}. While some models characterize the optical response through radiative and non-radiative decay rates, others incorporate additional physical mechanisms: pure dephasing processes for quantum mechanical interactions, and inhomogeneous broadening for structural imperfections and environmental variations. In this study, we experimentally measure the optical response of excitons in hBN-encapsulated monolayer TMDs at different temperatures and compare them to the different theoretical models suggested in the literature. This comparison enables us to evaluate which model more appropriately describes the excitonic optical response, with direct implications for predicting TMD behavior in complex structures and developing optoelectronic devices. Our results show that in high-quality heterostructures, quantum mechanical interactions in the form of pure dephasing plays a dominant role, which its contribution has been challenging to isolate experimentally in previous studies involving two-dimensional Fourier transform spectroscopy \cite{Moody2015IntrinsicDichalcogenides}. Furthermore, we find that accounting for an additional decay rate to the commonly used radiative and non-radiative rates is important for the accurate physical description of the excitonic response. \par

\begin{figure}[b]
    \captionsetup{justification=justified}
    \includegraphics[width=\columnwidth]{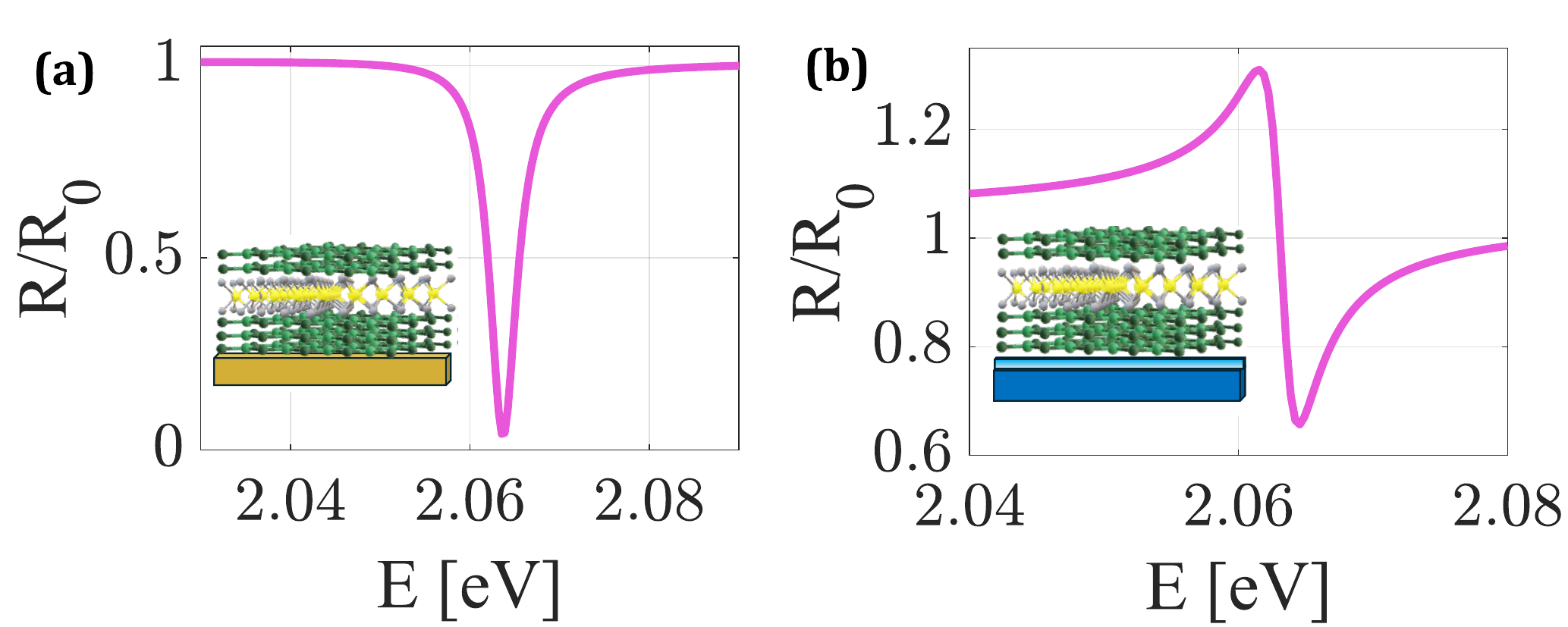}
    \caption{Exemplary reflection contrast spectra of commonly used heterostructures. Simulated reflection contrast for (a) hBN/TMD/hBN/gold mirror, and (b) hBN/TMD/hBN/$SiO_2$/$Si$. The insets illustrates the layered structures with the difference substrates.}
    \label{fig: simulation}
\end{figure}

\section{Physical Models}
Two experimental configurations are commonly employed, one where the hBN-encapsulated monolayer TMD is positioned on a metallic mirror that acts as a cavity designed to maximize excitonic absorption \cite{Epstein2020Near-unityCavity, Horng2020PerfectCrystal, Rogers2020CoherentExcitons}, which produces a characteristic Lorentzian lineshape in the reflection spectrum (figure \ref{fig: simulation}(a)); and the second where the heterostructure is placed on a $SiO_2$/$Si$ dielectric substrate, commonly yielding a Fano-shaped profile (figure \ref{fig: simulation}(b)) arising from the interaction between the excitonic response and multiple reflections within the layered structure \cite{Ugeda2014GiantSemiconductor, Epstein2020HighlySemiconductors}. To describe and understand these optical excitonic behaviors, several theoretical models have been proposed in the literature typically employing a frequency-dependent complex susceptibility $\chi(\omega)$, which characterizes the polarization response of the material to an applied electromagnetic field. These usually focus primarily on the first exciton resonance, i.e. the A exciton, while accounting for higher-energy contributions to susceptibility through a background term \cite{Epstein2020HighlySemiconductors, Li2014MeasurementE2, Epstein2020Near-unityCavity, Gershuni2024In-planeLoss}. A commonly used approach is an in-plane Lorentzian susceptibility model incorporating radiative and non-radiative decay rates, $\gamma_r$ and $\gamma_{nr}$, respectively, representing the rate at which the exciton population is lost by photon emission and various mechanisms of inelastic collisions, respectively \cite{Scuri2018LargeNitride, Li2014MeasurementE2, Back2018RealizationMoSe2, Shahmoon2017CooperativeArrays}:
\begin{align}\label{eq: General_Lorn}
\begin{split}
\chi_\perp = \chi_{bg} - \frac{c}{\omega_0d_0} \frac{\gamma_r}{\omega-\omega_0+i(\frac{\gamma_{nr}}{2})}
\end{split}
\end{align}
where $\chi_{bg}$ is the background susceptibility, $ c $ is the speed of light, $\omega_0$ is the exciton resonance frequency, and $d_0$ is the monolayer thickness. 

Other models describe the excitonic response in terms of the reflection coefficient \cite{Horng2020PerfectCrystal, Robert2018OpticalHeterostructures}: $r(\omega)=\frac{i\gamma_r}{\omega_0-\omega-i(\gamma_r+\gamma_{nr})}$. This approach relates the exciton properties to the macroscopic optical response of the material, facilitating comparisons with reflectance spectroscopy measurements \cite{Horng2020PerfectCrystal, Robert2018OpticalHeterostructures}. However, by expressing the reflection coefficient in terms of the TMD's conductivity $\sigma = -i\epsilon_0\omega d\chi$ and subsequently calculating the coefficient utilizing the transmission line model for a simple air/TMD interface (which will be elaborated upon later in this work), we can derive: $r(\omega) = \frac{-\sigma}{2\epsilon_0c+\sigma} = \frac{1}{2}\frac{i\gamma_r}{\omega_0-\omega-i\left(\frac{\gamma_r}{2}+\gamma_{nr}\right)}$, which is qualitatively equivalent to the Lorentzian susceptibility model of Eq. \ref{eq: General_Lorn} up to a normalization factor.

Scuri et al. \cite{Scuri2018LargeNitride}, initially present the Lorentzian susceptibility using the two decay rates model (Eq. \ref{eq: General_Lorn}), but then refine it by introducing a third, quantum pure dephasing rate, $\gamma_d$. This additional decay rate accounts for any type of elastic collision, such as exciton-exciton and exciton-phonon scattering, which do not alter the energy of the exciton but only in its phase. Here, $\gamma_d$ is introduced through the master equation with the proper Green's function, resulting in the overall reflection term:
\begin{align}\label{eq: model R with g_d}
    \begin{split}
        R(\omega) = \frac{\gamma_r^2}{2\tilde\gamma\gamma_T} \frac{1}{(\omega - \omega_0)^2/\tilde\gamma^2 +1}
    \end{split}
\end{align}
where $\gamma_T$ denotes the total decay rate ($\gamma_T = \gamma_r+\gamma_{nr}$), and $\tilde\gamma$ denotes the total dephasing rate ($\tilde\gamma = \gamma_d + \gamma_T/2$).

Epstein et al.\cite{Epstein2020HighlySemiconductors} similarly introduced an additional pure dephasing rate ($\gamma_d$) using a different modeling approach, replacing the Green's function with a quantum transfer-matrix-method (TMM) approach, which provides additional flexibility to simulate arbitrary configurations of van der Waals heterostructures, while still effectively treating the interaction with a quantum pure dephasing rate:
\begin{align}\label{eq: model 1}
    \begin{split}
    \chi_\perp = \chi_{bg} - \frac{c}{\omega_0d_0} \frac{\gamma_r}{\omega-\omega_0+i(\frac{\gamma_{nr}}{2} +\gamma_d)}
    \end{split}
\end{align}

Rogers et al.\cite{Rogers2020CoherentExcitons}, use the Lorentzian model of eq. \ref{eq: General_Lorn}, but introduce an additional decay rate, $\gamma_{ib}$, accounting for any source of inhomogeneous broadening through convolution with a Gaussian function:
\begin{align}\label{eq: model 3}
    \begin{split}
     R(\omega) = \frac{1}{\sqrt{2\pi\gamma_{ib}}} \int R_{\omega'_0} e^\frac{-(\omega_0 - \omega'_0)^2}{2\gamma_{ib}^2} d\omega'_0
    \end{split}
\end{align}

Using a phenomenological approach, Li et al. and others \cite{Li2021RefractiveDiselenide, Goncalves2018Plasmon-excitonInterfaces, Geisler2019Single-CrystallineTemperature} described a phenomenological Lorentzian complex susceptibility:
\begin{align}\label{eq: model 2}
    \begin{split}
    \chi = \chi_{bg} + \frac{f}{E_{0}^2-E^2-iE\Gamma}
    \end{split}
\end{align}
where $f$ represents the fitted oscillator strength and $\Gamma$ is the total linewidth of the exciton resonance, as measured at the FWHM.

Table \ref{tab:exciton_models} summarizes the different models of the excitonic optical response and their constituents, exhibiting a range of approaches that incorporate various physical mechanisms and decay rates to describe the excitonic response. The diversity of these models and the different amounts of physical constituents, together with their underlying physical mechanisms, raise the question of which model and corresponding physical mechanism most accurately represent the excitonic optical response in these materials. \par

\begin{table}
\caption{Summary of the physical models describing the optical response of excitons in monolayer TMDs.}
\label{tab:exciton_models}
\centering\Huge
\resizebox{\columnwidth}{!}{%
\begin{tabular}{|>{\centering\arraybackslash}m{4cm}|>{\centering\arraybackslash}m{14cm}|>{\centering\arraybackslash}m{5cm}|}
\hline
No. & Physical Model & Physical Parameters \\ [0.5ex] 
\hline
1 \textsuperscript{\normalsize\cite{Epstein2020HighlySemiconductors}} & \vspace{0.5cm} $\chi_\perp = \chi_{bg} - \frac{c}{\omega_0d_0} \frac{\gamma_r}{\omega-\omega_0+i(\frac{\gamma_{nr}}{2} +\gamma_d)} $ & 
{$\gamma_r$, $\gamma_{nr}$ , $\gamma_d$} \\ [3ex] 
\hline
2 \textsuperscript{\normalsize\cite{Li2021RefractiveDiselenide, Goncalves2018Plasmon-excitonInterfaces}} & \vspace{0.5cm} $\chi_\perp = \chi_{bg} + \frac{f}{E_0^2-E^2-iE\Gamma}$ & {$f$ ,$\gamma$} \\ [2ex] 
\hline
3 \textsuperscript{\normalsize\cite{Rogers2020CoherentExcitons}}&  \vspace{0.5cm} $\begin{array}{c}
\chi_\perp = \chi_{bg} - \frac{c}{\omega_0d_0} \frac{\gamma_r}{\omega-\omega_0+i(\frac{\gamma_{nr}}{2})} \\[2ex]
R(\omega) = \frac{1}{\sqrt{2\pi\gamma_{ib}}} \int R_{\omega'_0} e^\frac{-(\omega_0 - \omega'_0)^2}{2\gamma_{ib}^2} d\omega'_0
\end{array}$ & \vspace{0.5cm} {$\gamma_r$ ,$\gamma_{nr}$,$\gamma_{ib}$}  \\ [5ex] 
\hline
4 \textsuperscript{\normalsize\cite{Horng2020PerfectCrystal, Robert2018OpticalHeterostructures}} & \vspace{0.5cm} $\chi_\perp = \chi_{bg} - \frac{c}{\omega_0d_0} \frac{\gamma_r}{\omega-\omega_0+i(\frac{\gamma_{nr}}{2})}$ & {$\gamma_r$,$\gamma_{nr}$} \\ [2ex] 
\hline
\end{tabular}%
}
\end{table}

\section{Results}
\begin{figure}[t]
    \captionsetup{justification=justified}
    \includegraphics[width=1\columnwidth]{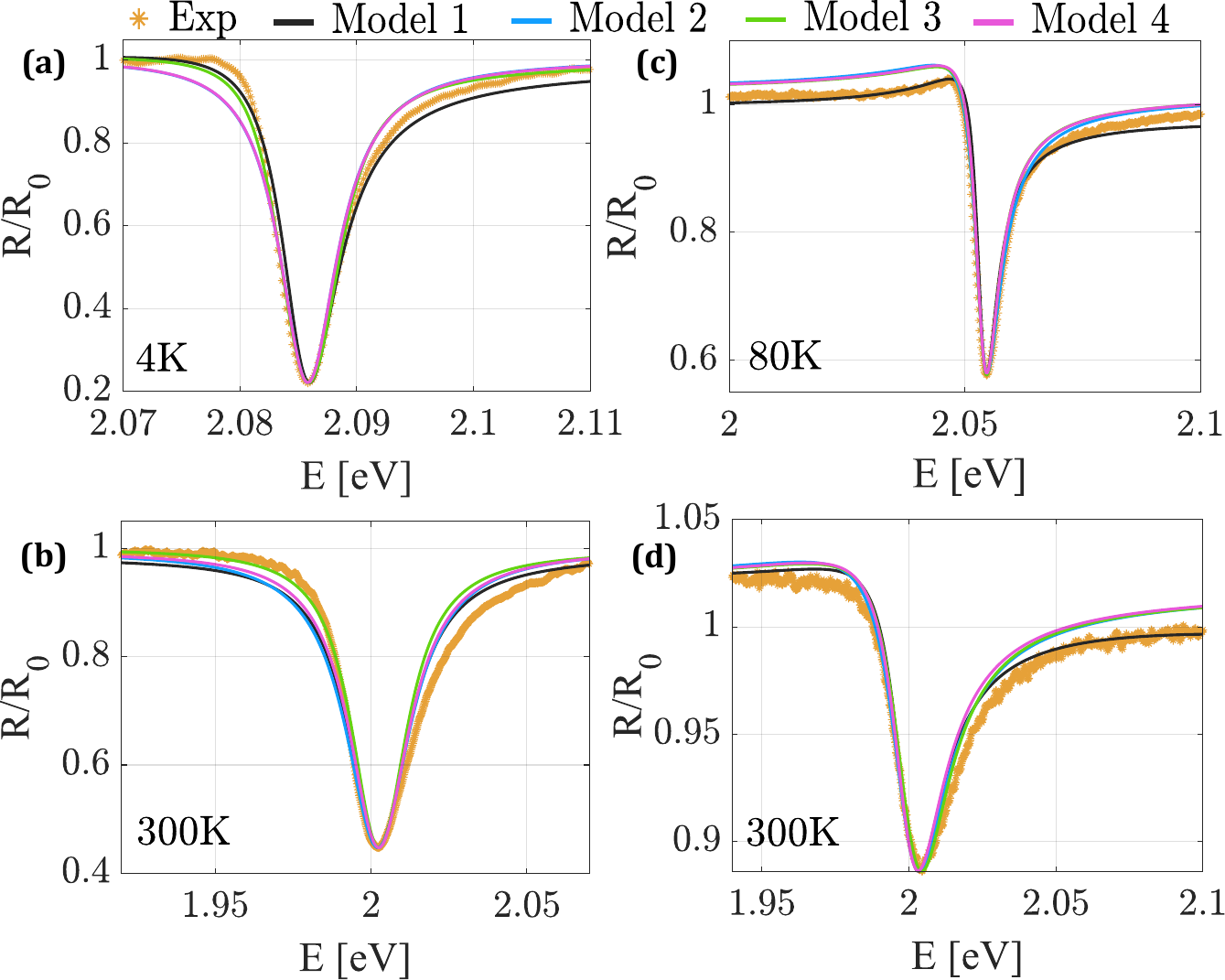}
    \caption{Experimental measurements and fitting of the reflection contrast at different temperatures. Measured spectra of sample A, consisting of 15 nm-hBN/monolayer $WS_2$/30 nm-hBN/gold mirror, at temperature of (a) $T=4K$, and (b) $T=300K$, respectively,  and of sample B, consisting of 60 nm-hBN/monolayer $WS_2$/40 nm-hBN/$SiO_2$/$Si$, at temperature of (c) $T=80K$, and (d) $T=300K$, respectively.  The curves show the models fitting to the measured results, and the expected exciton energy shifts and linewidth changes with temperature.}
    \label{fig: All Samples reflection}
\end{figure}

To evaluate the accuracy of the different models in predicting the optical response of TMD excitons, we fabricate two example samples for each configuration discussed in figure \ref{fig: simulation} and compare the experimentally measured reflection spectra with the theoretical predictions of each model. For both sample configurations, we have measured the temperature-dependent reflection at temperatures ranging from 5K to 300K, and we present the reflection contrast defined as $\frac {R}{R_0}$, where $R$ is the reflection from the full structure and $R_0$ is the reflection of the structure without the TMD \cite{Epstein2020HighlySemiconductors}. Exemplary experimental measurements of the reflection contrast for the two different sample configurations at two selected temperatures are presented in figure \ref{fig: All Samples reflection}, together with the fitting results from each one of the four models.\par 

It is important to note several fundamental behaviors in exciton linewidth; first the total linewidth increases with temperature owing to exciton-phonon interaction, as described by the phenomenological equation \cite{Moody2015IntrinsicDichalcogenides, Dey2016OpticalInteractions, Cadiz2017ExcitonicHeterostructures}
    \begin{align}\label{eq: Temperature Vs. linewidth}
        \begin{split}
            \gamma_i(T)
            =\gamma_i(0)+c_1T+\frac{c_2}{e^{\frac{\Omega}{k_BT}}-1}
        \end{split}
    \end{align}
where $\gamma_i(0)$ represents a temperature-independent offset, $c_1$ accounts for the linear increase due to acoustic phonon interactions, $c_2$ denotes the strength of the optical phonon coupling, and $\Omega$ is the average energy of the involved phonons. This model effectively captures the fundamental aspects of the exciton-phonon interactions that contribute to linewidth broadening and can thus be used as a guiding rule for the temperature-dependent broadening \cite{Moody2015IntrinsicDichalcogenides, Epstein2020HighlySemiconductors, RudinTemperature-dependentSemiconductors, Scuri2018LargeNitride}. The linear term in Eq. \ref{eq: Temperature Vs. linewidth} describes the exciton interaction with acoustic phonons, whereas the last term accounts for interactions with longitudinal optical phonons. Second, the radiative decay rate is known not to be temperature-dependent and is affected by the exciton's environment through the Purcell effect \cite{Epstein2020HighlySemiconductors, Scuri2018LargeNitride, Horng2019EngineeringSemiconductors, Fang2019ControlHeterostructures}. We will use both these temperature dependencies to verify and validate our results. \par

For the comparison of the results to the models presented in TABLE \ref{tab:exciton_models}, we use an analytical calculation based on the transmission line model (TLM), which is a versatile approach for solving wave propagation problems in layered structures \cite{Eini2022Valley-polarizedFrequencies, Eini2022Valley-polarizedSemiconductors, Kats2DMaterials}, where each layer is represented by its characteristic admittance, $Y_i = \frac{\omega\epsilon_0\epsilon_i}{k_i}$, where $k_i,\epsilon_i$ are the wavevector and permittivity of the i-th layer, respectively. The reflection coefficient for the entire layered structure is then obtained from:
\begin{align}\label{eq: admittance}
        r &= \frac{Y_2-Y_{air}}{Y_2+Y_{air}}
\end{align}
\par
where $Y_{air}$ is the characteristic admittance of air and $Y_2$ is the characteristic admittance of the structure.

\begin{figure}[b]
    \captionsetup{justification=justified}
    \includegraphics[width=\columnwidth]{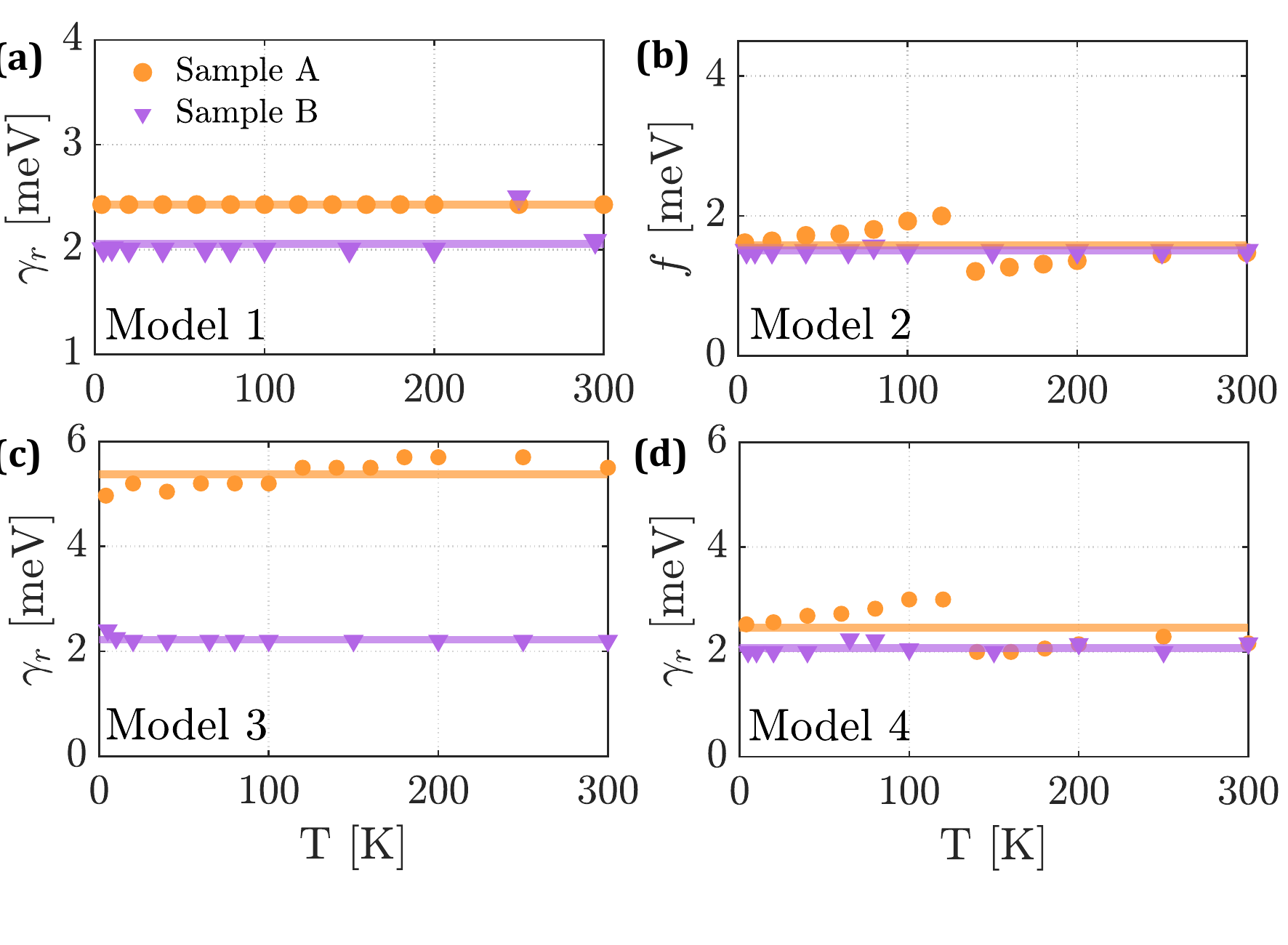}
    \caption{Temperature-dependent radiative response of samples A and B. (a),(c),(d) Radiative decay rate for models 1, 3, and 4, respectively. (b) Oscillator strength for model 2. The radiative decay rates and oscillator strengths exhibit a weak and even temperature-independent behavior.}
    \label{fig: g_r}
\end{figure}

 The four models show varying degrees of success in fitting the experimental data: As shown in figure \ref{fig: All Samples reflection}(a) and (b), the reflection contrast measurements display the expected Lorentzian lineshape with a pronounced absorption dip, consistent with the predicted behavior shown in figure \ref{fig: simulation}(a). Sample B measurements (figure \ref{fig: All Samples reflection}(c) and (d)) exhibit the characteristic Fano-shaped profile as predicted in figure \ref{fig: simulation}(b). Both samples demonstrate clear spectral signatures that align with their theoretical predictions, enabling a detailed analysis of their excitonic properties. All samples show the expected temperature-dependent behavior of exciton energy shifts and linewidth narrowing at lower temperatures, as previously observed \cite{ Scuri2018LargeNitride, Horng2020PerfectCrystal, Epstein2020Near-unityCavity, Epstein2020HighlySemiconductors}.\par

\begin{figure}[t]
    \captionsetup{justification=justified}
    \includegraphics[width=\columnwidth]{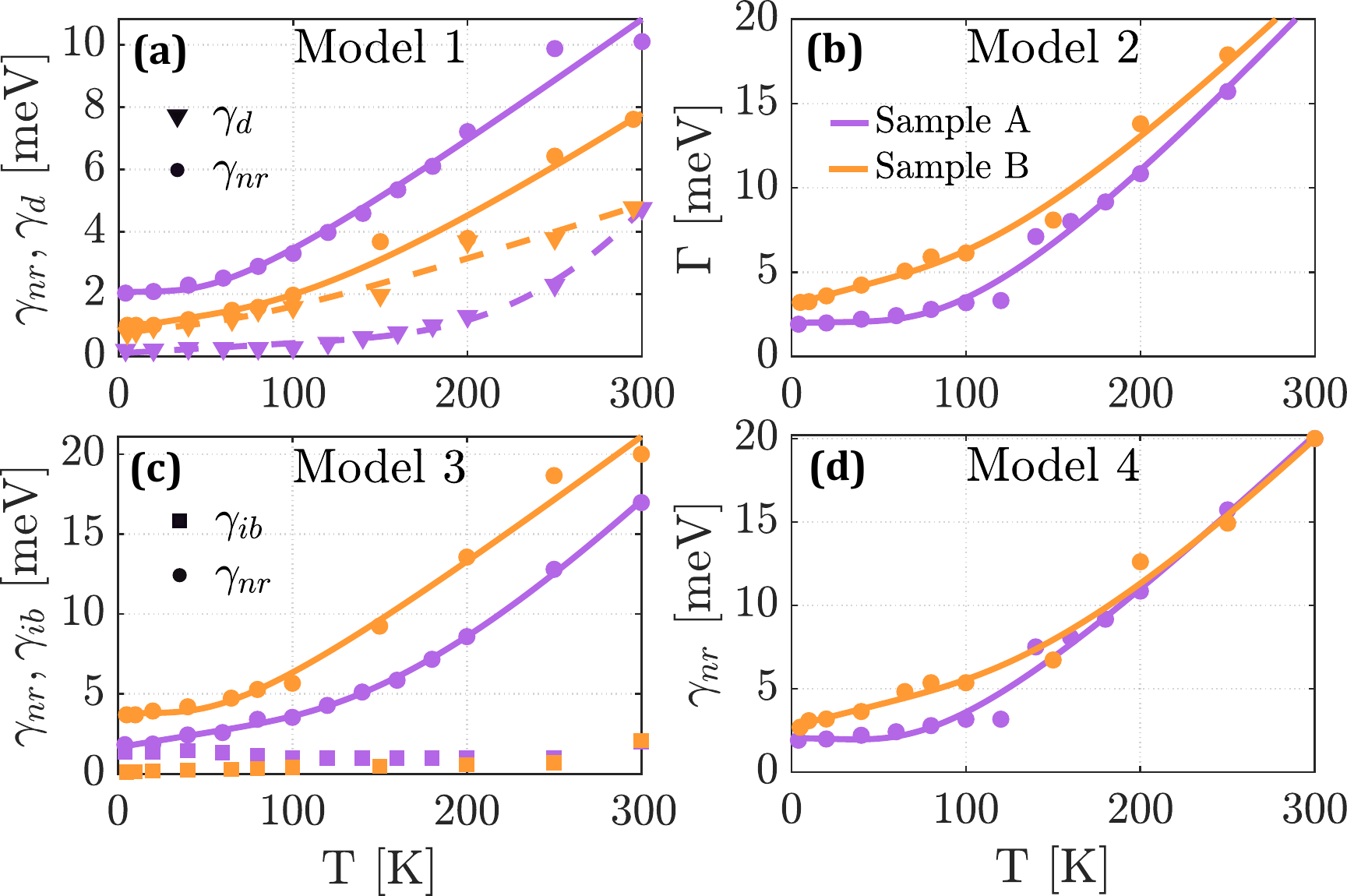}
    \caption{Temperature-dependent non-radiative, pure dephasing and inhomogeneous broadening response of samples A and B. (a) Non-radiative and pure dephasing decay rates for model 1. (b) Total linewidth for model 2. (c) Non-radiative and inhomogeneous broadening decay rates for model 3. (d) Non-radiative decay rate for model 4. The extracted parameters show the expected temperature dependence, with $\gamma_{nr}$, $\gamma_d$, and $\Gamma$ following the phenomenological model of Eq. \ref{eq: Temperature Vs. linewidth}, while $\gamma_{ib}$ remains constant with temperature.}
    \label{fig: All params}
\end{figure}

 First, we examine the temperature-dependent behavior of the extracted decay rates to validate the physical consistency of our analysis. Figure \ref{fig: g_r}(a-d) present the temperature dependencies of $\gamma_r$ and $f$, extracted from fitting for all samples and models. In models 1, 3, and 4, $\gamma_r$ exhibits small variations with temperature (figure \ref{fig: g_r}(a,c,d)), which agrees well with previous observations \cite{Scuri2018LargeNitride, Epstein2020HighlySemiconductors, Selig2016ExcitonicDichalcogenides}, as it is only affected by the exciton's environment through the Purcell effect \cite{Horng2019EngineeringSemiconductors, Epstein2020Near-unityCavity, Fang2019ControlHeterostructures}. The values of the radiative decay rates also agree with previous reports \cite{Cadiz2017ExcitonicHeterostructures, Scuri2018LargeNitride, Epstein2020HighlySemiconductors}. Similarly, $f$ also remains approximately constant with temperatures (figure \ref{fig: g_r}(b)), as it is directly related to the radiative decay rate \cite{Epstein2020Near-unityCavity} .\par

Figure \ref{fig: All params}(a-d) exhibits the temperature-dependent behavior of $\gamma_{nr}$, $\gamma_d$ and $\Gamma$, demonstrating a clear temperature response following the expected phenomenological model of Eq. \ref{eq: Temperature Vs. linewidth}. The temperature-related changes in these decay rates revealed two distinct patterns: at lower temperatures (10–100 K), we note a linear increase, suggesting the predominance of acoustic phonon interactions \cite{Selig2016ExcitonicDichalcogenides, Mueller2018ExcitonSemiconductors}, while above 100K, the samples display a superlinear relationship, indicating a growing influence of optical phonons. These results agree well with previous reports \cite{Rogers2020CoherentExcitons, Horng2020PerfectCrystal, Scuri2018LargeNitride, Selig2016ExcitonicDichalcogenides}. The inhomogeneous broadening parameter of model 3 $\gamma_{ib}$ exhibits minimal variation with temperature (figure \ref{fig: All params}(c)), which is consistent as well with previous studies  \cite{Selig2016ExcitonicDichalcogenides, Dey2016OpticalInteractions}. \par

\begin{figure}[b]
    \captionsetup{justification=justified}
    \includegraphics[width=\columnwidth]{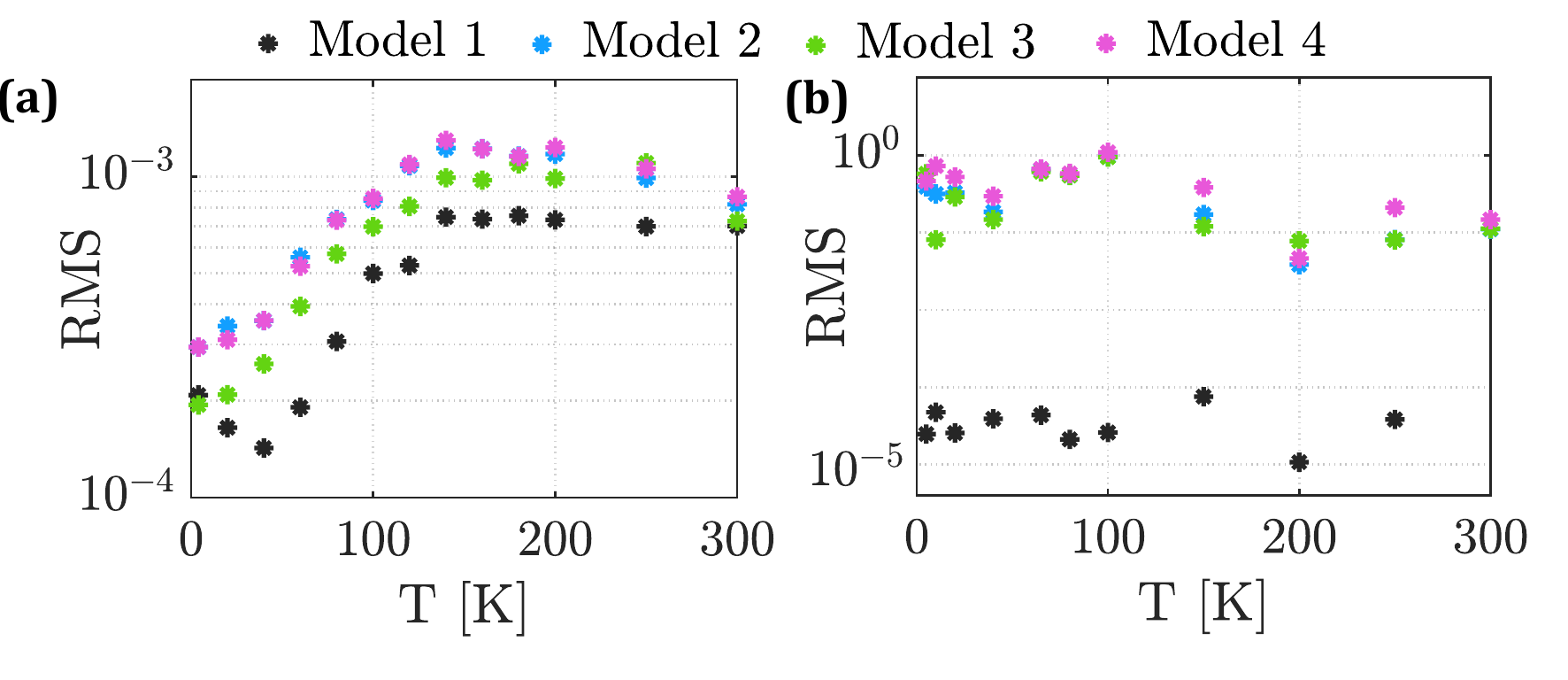}
    \caption{The RMS fitting figure-of-merit for (a) Sample A and (b) Sample B, showing that model 1, which incorporates the pure dephasing rate, provides the most accurate fit to the experimental data. Moreover, it can be seen that the general use of models incorporating three decay rates (models 1 and 3), qualitatively provides better results than those accounting only for two decay rates (models 2 and 4).}
    \label{fig: Fitting}
\end{figure}

Finally, we examine the accuracy of the different models fits to the experimental results by defining the root mean square (RMS) error of the fit as the figure-of-merit. Figure \ref{fig: Fitting} presents the calculated RMS for all the models' fits for both samples, showing two distinct overall behaviors; first, it is clearly seen that the best fit for both samples is obtained for model 1, which accounts for the quantum pure dephasing rate. Second, it can be seen that the general use of models incorporating three decay rates (models 1 and 3), qualitatively provides better results than those accounting only for two decay rates (models 2 and 4). Specifically, sample A (figure \ref{fig: Fitting}(a)) exhibits consistently lower RMS values for Model 1 compared to other models across all temperatures.  In sample B (figure \ref{fig: Fitting}(b)), Model 1's RMS values are nearly an order of magnitude below models 2, 3, and 4, which show less pronounced differences between themselves.

\section{Discussion}
From the above results, it is clear that accounting for an additional decay rate describing another physical phenomenon, beyond the common radiative and non-radiative rates, better grasp the physical response of the system. Moreover, it seems that the quantum pure dephasing rate, in particular, plays an important role in our samples. We believe that this stems for the fact that both of our samples incorporate the use of high-quality growth TMD crystals (see SI) with very low defect density, which together with the contribution of the high-quality hBN encapsulation (see SI) \cite{Liu2018SingleNitride} significantly reduces structural disorder and impurities in the samples, thereby reducing the contribution of inhomogeneous broadening. This is in contrast with the work of Rogers et al.\cite{Rogers2020CoherentExcitons} that have used commercially grown TMD crystals, which may explain their choice in adding an inhomogeneous broadening decay rate. From the adequate treatment of the pure dephasing rate through the quantum TMM approach, under these optimized conditions it seems that the exciton linewidth broadening relates to quantum mechanical interactions affecting the phase of the exciton wave function without changing its energy, such as elastic scattering from exciton-exciton or exciton-phonon interaction. This finding is especially significant given that previous attempts to experimentally isolate the pure dephasing contribution to the linewidth, from temperature-dependent studies of exciton dynamics via two-dimensional Fourier transform spectroscopy, were unable to directly measure the pure dephasing rate owing to the dominant influence of population relaxation in the sample\cite{Moody2015IntrinsicDichalcogenides}.  

In conclusion, through experimental spectroscopy of high-quality TMD-based van der Waals heterostructures, we have been able to evaluate the commonly used theoretical models describing the optical response of excitons in monolayer TMDs. We found that incorporating pure dephasing (Model 1) as a third decay parameter provides the best fitting across our experimental conditions. These results advance our understanding of the fundamental physical behavior in these systems and their potential application in optoelectronic and valleytronic devices.\par

\section{ACKNOWLEDGMENTS}
I.E. acknowledges the Israeli Science Foundation personal grant number 865/24. S.T. acknowledges support from Applied Materials Inc (materials development) and Lawrence Semiconductors (metrology), J.H.E. and T.P. appreciate the support for hBN crystal growth from the Office of Naval Research, award number N00014-21-1-2939.
 
\bibliography{references}

\end{document}